----------------------------- Start of body part 2

\documentstyle[12pt]{article}
\newcommand{\beq}{\begin{equation}}
\newcommand{\eeq}{\end{equation}}
\newcommand{\beqa}{\begin{eqnarray}}
\newcommand{\eeqa}{\end{eqnarray}}
\newcommand{\bc}{\begin{center}}
\newcommand{\ec}{\end{center}}
\newcommand{\initi}{\mid\Psi_{i}>}
\newcommand{\inito}{\mid\Psi^{(1)}_{i}>}
\newcommand{\initt}{\mid\Psi^{(2)}_{i}>}
\newcommand{\final}{<\Psi_{f}\mid}
\newcommand{\finalo}{<\Psi^{(1)}_{f}\mid}
\newcommand{\finalt}{<\Psi^{(2)}_{f}\mid}
\newcommand{\initii}{<\Psi_{i}\mid}
\newcommand{\initiio}{<\Psi^{(1)}_{i}\mid}
\newcommand{\initiit}{<\Psi^{(2)}_{i}\mid}
\newcommand{\inte}{\int_{0}^{T} \hat{H} dt}
\newcommand{\produ}{\lim_{{\epsilon\rightarrow 0}_{N\rightarrow
\infty}}\prod_{j=1}^{N}{\it{U}}(t_{j},t_{j-1})}

\newcommand{\infek}{\hat{U}(t_{k}, t_{k-1})}

\newcommand{\infj}{\hat{\cal{U}}(t_{j}, t_{j-1})}
\newcommand{\ufk}{\hat{\cal{U}}(t_{f}, t_{{k}_{1}})}
\newcommand{\ufkk}{\hat{\cal{U}}(t_{f}, t_{{k}_{2}})}
\newcommand{\ukkk}{\hat{\cal{U}}(t_{{k}_{1}},t_{{k}_{2}})}
\newcommand{\uki}{\hat{\cal{U}}(t_{{k}_{1}}, t_{0})}
\newcommand{\ukki}{\hat{\cal{U}}(t_{{k}_{2}}, t_{0})}
\newcommand{\sumacc}{(u^{*}\hat{\eta}^{\dagger}
+ v^{*}\hat{\rho}^{\dagger})}
\newcommand{\suma}{(u\hat{\eta} + v\hat{\rho})}
\newcommand{\limite}{\lim_{{\epsilon \rightarrow 0}
_{N \rightarrow \infty}}}

\title{Mean Field Approximation to two-subsystems: Examples}
\author{S. Cruz-Barrios\\
Departamento de Fisica Aplicada \\
E.U.I.T.A, Ctra de Utrera Km 1. \\
Universidad de Sevilla}

\begin{document}

\maketitle

\begin{abstract}
Using the Funtional Integrals Formulation  is developed a
self-consistent mean field expansion  to evolution operators
of a system composed by two subsystems. This is a general expansion
and can be generalized for more of two subsytems, which
can be as system composed by fermions and bososns with an
interactin between their ($\sigma$-model,$\sigma$-$\omega$
model or maser model )   as a fermionic system with two
or more different interaction between fermions, (NJL model).

\end{abstract}

\section{Introduction}

In a recent papers\cite{cn} we  shown a scheme
to  treat field theoretical Lagrangians in the
same base of well known nonrelativistic many body
techniques. Fermions and bosons were treated quantum
mecanically in a symmetric way  obtaining results
for the mean field approximations. This same approximation,
when used in Nambu Jonas Lasino Model, reveals interesting
propreties in terms of currentes language on
nonrelativistic many body physics. Several questions  such that,
what does happen in NJL model after mean field approximation?
or what do we can learn from this approximation ?
or in systems such that fermions and bosons
are treated quantum mecanically in a symmetric way, How is the RPA?
or How are the correlations in these systems? Is it possible go
beyond  RPA?,
are importants
and the answer envolve  a self-consistent mean-field expansion.
As in the NJL case as in  fermions plus bosons case
or in nonrelativistic case such that, Maser Dick model,  the system
to be treated contain two (or more) subsystem or interaction.
In NJL model, in the more simple version,
 we will have a fermions system with a scalar and
pseudo vector interaction, in fernions and bosons case, when treated
quantum mecanically  in a symmetric way  or in maser model,
(spin system, fermions, and electromagnetic field, bosons),
we have  two differents subsystem explicitly.

The question is,
How do develope for two or more subsystem
 a self-consistent mean-field expansion? \\

In\cite{cp} we  developed a mean-field expansion for the
many-fermions initial condition problem on the basis of the flexible
funtional integral representation. To each order in the expansion,
one considers the transition amplitudes from the prescribed
initial state to the final states to which it self-consistently
evolves. With this considerations was possible to obtain a mean-field
expansion to evolution operator for the many-body system with a
two-body interaction and show as appear the two-body correlation in
fourth order expansion.

Using the same idea and formalism that in\cite{cp} is possible
to obtain a self-consistent mean-field expansion to
evolution operator of a system composed with two or more
subsystem. This will be a general expansion and could be
applied as in NJL case as in any  general system composed by two
or more subsystem.  One of the reason to look for an expantion
to the evolution
operator  is  have the oportunity  to  analise the dynamical
proprieties as on nonrelativistic as in  relativistic systems
with two or more subsystems. \\

The paper is organized as follows. In section 2 we set up the problem
of obtaining an expansion for evolution operators $\hat{U}(t,0)$
by considering the functional representation of the transition matrix
elements $\final \hat{U}(t,0)\initi $ with the subsidiary
"self-consistent" condition that $\final$ in given as the time-evolved
initial state $\initi$ to the desired order approximation. The
self-consistent mean-field expansion is obtained in section 3, where
terms up to and including the fourth order are worked out. There it is
shown explicitly that correlation effects first  contribute to
the dynamics in fourth order. The $\sigma-model$ system is used as
a symple and didatic example, illustrating the self-consistent
mean-field approximation obtained with this formalism in section 4.
Finally, conclutions and comments are given in section 5.

\section{Description of the System and Functional Integral
Representation}

As said in the introduction, the system of interesting
is  composed by two (or more) subsystem, such as
fermions (nucleons) plus bosons (mesons) particles
(as in maser model
or $\sigma$ model, etc),  N-fermions particles with
two or more differents interaction, (as in the NJL case),
or any general system for which will be possible to distinguish
between to or more subsystem. So, the more general system with this
 chararteristic can be described by a
hamiltonian $\hat{H}$,  composed by
a subsystem, $(1) \equiv \hat{H}_{1}$
and $(2)\equiv \hat{H}_{2}$, with an interaction
between their, $H_{12}$ such that, we go
to separete by convenience of the following way:
\beq
\hat{H} = \hat{H}^{1}_{1} + \hat{H}^{1}_{2} + \hat{H}_{int}
\eeq
with: \\
{\bf{a}}- $\hat{H}^{1}_{1}$, corresponding at the Kinetic
contribution of the system one (1) and
it can be represented, in the more general form as: \\
\[
{\hat{H}}^{1}_{1} = K_{1}\hat{\rho} \equiv
\sum_{\alpha \beta} K_{{1}_{\alpha\beta}}a^{\dagger}_{\alpha}a_{\beta}
\]
{\bf{b}}- $\hat{H}^{1}_{2}$, corresponding at the kinetic contribution
of the second $(2)$ system  and one  represent of general way by :
\[
{\hat{H}}^{1}_{2} = K_{2}\hat{\eta} \equiv
\sum_{\alpha \beta} K_{{1}_{\alpha\beta}}b^{\dagger}_{\alpha}b_{\beta}
\]
{\bf{c}}- $\hat{H}_{int}$, corresponding at two-body
interaction from systems $(1)$, $(2)$  and  the interaction between
their, and can be repesented, in a general way
as:
\[
{\hat{H}}_{int} =  \hat{H}^{2}_{1} + \hat{H}^{2}_{2} + \hat{H}_{12}
\]
with:
\[
{\hat{H}}^{2}_{1}  =  V_{1}\hat{\Omega}_{1} \]
\[ {\hat{H}}^{2}_{2}  =  V_{2}\hat{\Omega}_{2} \]
\[ {\hat{H}}_{12}     =  K_{12}\hat{\rho}\hat{\eta}^{\dagger} +
K^{*}_{12}\hat{\rho}^{\dagger}\hat{\eta} \]
where, $V_{i}$ is responsible for two-body interactions
$\Omega_{i}$ is the two-body operators, and $K_{12}$ is
the interaction between (1)-(2) subsystem. \\

So, the system dsecrided by hamiltonian, $ H $ eq.(1) are representing
any two-subsystem, $(1) \equiv H_{1}$ and $(2) \equiv H_{2}$
with one and two body operators in both subsystems and
one interaction, $(H_{12})$ between their verifing:
\[
{[{\hat{H}}_{1}, {\hat{H}}_{2}] } = 0 \]
\[ {[{\hat{H}}_{1}, {\hat{H}}_{12}] } \neq  0  \]
\[ {[{\hat{H}}_{2}, {\hat{H}}_{12}] }  \neq 0  \]

\vspace{2.5cm}

The time evolution of the system is given by the
operator $\hat{U}(t,0)= exp\{\inte\}$,
and $\hat{H}$ is
representing the hamiltonian  of the total system
$\hat{H} = \hat{H}^{1}_{1} + \hat{H}^{1}_{2} + \hat{H}_{int}$.
The evolution operator $\hat{U(t,0)}$ can be written as an
infinitesimal temporal product of evolution operators,
$U(T,0) = \produ $,
with,  ${\it{U}}(t_{j}, t_{j-1}) = 1- i\epsilon \hat{H}$.
Using the same prescription of the reference\cite{cp}(and other
references therein),
the contribution of the interaction term
$\hat{H}_{int}$ in the infinitesimal evolution operator,
can be written as:
\beqa
i\epsilon \hat{H}_{int}& = & \bar{N} \int d \sigma d \sigma^{*}
exp\{i\frac{\epsilon}{2} \mid \sigma \mid^{2}\}
\nonumber \\
                      &    & - \frac{\epsilon^{2}}{2q}(\sigma \hat{H}_{int}
\sigma^{*} +\sigma^{*}\hat{H}_{int} \sigma)
\eeqa
with $\sigma$ and $\sigma^{*}$ a $c-numbers$ matrix and $q$ a parameters.
The new expression to  $i\epsilon \hat{H}_{int}$
is now a gaussian funtional integrals, consequently any combinations of
lineal terms in
 $\sigma$ and $\sigma^{*}$
have a null contribution in total gaussian integrals. We should use
this fact to introduce a mixing terms between $\sigma$, $\hat\rho$,
$\hat{\eta}$ and $\sigma^{*}$ which should go to be  important
in the mean-field approximation description into  two-subsystem.

With this considerations about gaussian integrals the expression (2)
is equivalent (in exact calculation) to :
\beqa
i\epsilon \hat{H}_{int} & = & \bar{N} \int d \sigma d \sigma^{*}
exp \{ i\frac{\epsilon}{2}
\mid\sigma \mid^{2} \}  \nonumber\\
                       &   & \{ i \epsilon \sigma(u^{*}
\hat{\eta}^{\dagger} +
v^{*}\hat{\rho}^{\dagger}) + i\epsilon\sigma^{*}
(u\hat{\eta} + v\hat{\rho}) +  \nonumber \\
                       &   & \frac{\epsilon^{2}}{2q} (\sigma\hat{H}_{int}
\sigma^{*} +\sigma^{*}\hat{H}_{int} \sigma) \}
\eeqa

The introducion of lineal terms is completly arbritary and we have
total liberty to choose its, of this way in case of more of two
differents
operators or more of two differents interaction in the system,
we can introduce as numbers of operators as differents
kind of interaction we have in the system.

The various terms appearing in above expressions for
$i\epsilon\hat{H}_{int}$
are written out explicitly as in ref.\cite{cp}:
\[
{\mid}\sigma\mid^{2}  \equiv  \sum_{\alpha \beta}\sigma_{\alpha\beta}
\sigma^{*}_{\beta \alpha} \]
\[ {\sigma}(u^{*}\hat{\eta}^{\dagger} + v^{*}\hat{\rho}^{\dagger} )
   \equiv  \sum_{\alpha\beta\gamma\delta}
\sigma_{\alpha\beta}(u^{*}_{\beta\alpha\delta\gamma}
\hat{\eta}^{\dagger}_{\delta\gamma}
+ v^{*}_{\beta\alpha,\delta\gamma}\hat{\rho}^{\dagger}_{\delta\gamma})
\]
\[ {\sigma}^{*}(u\hat{\eta} + v\hat{\rho})
  \equiv \sum_{\alpha\beta\gamma\delta}
\sigma^{*}_{\beta\alpha}(u_{\alpha\beta,\gamma\delta}
\hat{\eta}_{\gamma\delta} +
v_{\alpha\beta,\gamma\delta}\hat{\rho}_{\gamma\delta}) \]
\[ {\sigma}\hat{H}_{int}\sigma^{*} \equiv
\sum_{\alpha\beta\gamma\delta\lambda\mu}
\sigma_{\alpha\beta} \left( \left[ K_{{12}_{\alpha\beta,\gamma\delta}}
\hat{\rho}_{\gamma\delta}
\hat{\eta}^{\dagger}_{\mu\lambda} +
 K^{*}_{{12}_{\alpha\beta,\gamma\delta}}
\hat{\eta}_{a\gamma\delta}
\hat{\rho}^{\dagger}_{\mu\lambda}\right] \right. + \]
\[ {\left. \;\;\;\;\;\;\;\;\;\;\;\;\;\;\;\;\;\;\;
V_{{1}_{\alpha\gamma\beta\delta}}
\hat{\Omega}_{{1}_{\gamma\delta\lambda\mu}}
+
V_{{2}_{\alpha\gamma\beta\delta}}
\hat{\Omega}_{{2}_{\gamma\delta\lambda\mu}} \right )}
 \sigma^{*}_{\mu\lambda} \]

\vspace{2.5cm}

The infinitesimal evolution operator
$U(t_{k}, t_{k-1}) = 1 -i\epsilon \hat{H} $, can be written
as:

\beqa
\infek & = & \bar{N}\int d\sigma_{k} d\sigma^{*}_{k}
exp\{ i\frac{\epsilon}{2}
\mid \sigma_{k}\mid^{2} \} \nonumber \\
      &   & \left \{1 - i\epsilon \hat{H}^{1}_{1} - i\epsilon
\hat{H}^{1}_{2}- \right. \nonumber \\
      &   & -i\epsilon \left[ \sigma_{k}(v^{*}\hat{\rho}^{\dagger} +
u^{*}\hat{\eta}^{\dagger})_{k} + \sigma^{*}_{k}(v\hat{\rho}
+ u\hat{\eta})_{k}\right]  \nonumber \\
      &   & - \left. \frac{\epsilon^{2}}{2q}
\left(\sigma_{k} \hat{H}_{int} \sigma^{*}_{k} +
\sigma^{*}_{k} \hat{H}_{int} \sigma _{k}\right) \right \}
\eeqa

and the total evolution operator is:

\beqa
U(T,0) = & \bar{N} \int D[\sigma] D[\sigma^{*}] \limite \tau
\prod^{N}_{k=1}  exp \{ i\frac{\epsilon}{2} \sum_{k}
\mid \sigma_{k} \mid^{2} \} \nonumber \\
         & \{ 1 - i\epsilon \hat{H}^{1}_{1} - i\epsilon
\hat{H}^{1}_{2} - \nonumber \\
         & - i\epsilon \sigma_{k} ( v^{*}\hat{\rho}^{\dagger}
+ u^{*}\hat{\eta}^{\dagger} )_{k}
-i\epsilon \sigma^{*}_{k}(v\hat{\rho} + u\hat{\eta})_{k}
\nonumber \\
         & - \frac{\epsilon^{2}}{2 q}(\sigma_{k}
\hat{H}_{int}\sigma^{*}_{k} + \sigma^{*}_{k}\hat{H}_{int}\sigma_{k}) \}
\eeqa

where
\bc
\[
{\int} {\it{D}}[\sigma]{\it{D}}[\sigma^{*}]  = \limite
\tau \int \prod^{N}_{k=1} d \sigma_{k} \prod^{N}_{k=1} d\sigma^{*}_{k}
\]
\ec

with \\
\bc
\[d \sigma_{k} =  \prod _{\alpha \beta} d\sigma_{\alpha \beta} \] \\
\ec
and \\
\bc
\[ {\bar{\it{N}}} = \left[ \limite
\prod^{N}_{k=1} \int d\sigma_{k} d\sigma^{*}_{k} exp \{i \frac{\epsilon}{2}
\sum_{k}\mid \sigma_{k}\mid^{2} \} \right]^{-1} \]
\ec
the normalization constant.

Theses relations define the evolution operator as a funtional
integral. \\ \\

To obtain the self-consistent mean field expansion, we will use
the same argument and idea that in reference\cite{cp}, where
we studied the transition amplitude,  $\final U(T,0) \initi $
with initial and final states $ \initi , \final $ respectively.
The situation here is similar, with the difference that
the initial and final state are
a composition of the subsystem $1$ and $2$, consequently we have
\beq
\final U(T,0) \initi = \finalo \finalt U(T,0) \initt \inito
\eeq
where $\finalo \finalt$ and $\inito \initt$, is representing the
initial and final state of the subsystem $1$ and $2$, respectively. \\

Replacing (5) in (6), the transition amplitude is now represented
as a funtional integral of the following way:
\beq
\final U(T,0) \initi = \int {\it{D}}[\sigma] {\it{D}}[\sigma^{*}]
e^{i S[\sigma, \sigma^{*}]}, \\
\eeq
where,
\[
S[\sigma, \sigma^{*}] =  \frac{\epsilon}{2} \sum^{N}_{k=1}
\mid\sigma_{k}\mid^{2} - iln \final {\it{U}}(T,0) \initi \] \\
can be identifed as an action governing the sought time evolution.
Furthermore,

\beqa
{\it{U}}(t,0) = & \limite \tau \prod^{N}_{k=1}[&
1 - i\epsilon \hat{H}^{1}_{1} - i\epsilon \hat{H}^{1}_{2} - \nonumber \\
                &                              &
-i\epsilon \sigma_{k}(u^{*}\hat{\eta}^{\dagger} +
v^{*}\hat{\rho}^{\dagger})_{k}
-i\epsilon \sigma^{*}_{k}(u\hat{\eta} + v\hat{\rho})_{k}
\nonumber \\
                &                              &
 -\frac{\epsilon^{2}}{2q} (\sigma _{k}\hat{H}_{int}\sigma^{*}_{k} +
\sigma^{*}_{k}\hat{H}_{int}\sigma_{k})]
\eeqa
is a product of infinitesimal time displecements involving the
auxiliaries fields, $\sigma(t')$ and $\sigma^{*}(t')$, $\tau$,  stading
for the time-ordering symbol.

\subsection{Stationary Phase  Approximation}

In order to do the stationary phase expansion, we write the auxiliary
field $\sigma$ and $\sigma^{*}$ as its stationary value $\bar{\sigma}$
($\bar{\sigma}^{*}$ )
plus a fluctuation parte $\xi$ ($\xi^{*}$),
\bc
\[
{\sigma_{k}}     = {\bar{\sigma}}_{k} + \xi_{k} \]
\[
{\sigma}^{*}_{k} =  {\bar{\sigma}^{*}}_{k} + \xi^{*}_{k} \]
\ec
so that the stationary phase is found from the conditions
\bc
\[ {\left(
\frac{\delta S}{\delta \xi_{k}} \right)}_{\xi =\xi^{*} = 0}  =
\left(
\frac{\delta S}{\delta \xi^{*}_{k}} \right)_{\xi = \xi^{*} = 0} = {0}
\]
\ec

The $\limite $
together with the SPA condition, lead in a straighforward way to
 $\bar{\sigma}_{k}$ and $\bar{\sigma}^{*}_{k}$:
\beq
\bar{\sigma}^{*}_{k} =
2 \limite \frac{\final \tau \prod^{N}_{j=k+1}\infj
\sumacc_{k} \prod^{k-1}_{j=1}\infj \initi}
{\final \tau \prod ^{N}_{j=1} \infj \initi} \\
\eeq
and:

\beq
\bar{\sigma}_{k} =
2 \limite \frac{\final \tau \prod^{N}_{j=k+1} \infj \suma_{k}
\prod^{k-1}_{j=1} \infj \initi}
{\final \tau \prod^{N}_{j=1} \infj \initi} \\
\eeq
with $\infj$ the infinitesimal evolution operator of the total
system in the SPA, given by
\beq
\infj = 1 - \epsilon \hat{h}_{j}. \\
\eeq

$\hat{h}_{j}$ is the hamiltonian
in this aproximation and represent the
hamiltonian of the two subsystem, ie,

\[ {\hat{h}}_{j}  = \hat{h}_{{1}_{j}} + \hat{h}_{{2}_{j}} \]
with:
\beqa
\hat{h}_{{1}_{j}} & = &  \hat{H}^{1}_{1} + \bar{\sigma}_{j}
v^{*}\hat{\rho}^{\dagger} + \bar{\sigma}^{*}_{j}v\hat{\rho} \;\;\;\;\;\;\;
(a) \nonumber \\
\hat{h}_{{2}_{j}} & = &  \hat{H}^{1}_{2} +
\bar{\sigma}_{j}u^{*}\hat{\eta}^{\dagger}
+ \bar{\sigma}^{*}_{j}u\hat{\eta} \;\;\;\;\;\;\; (b) \nonumber \\
\eeqa

We  observe, in (9) and (10), the  explicity dependence into
stationary phase, $\bar{\sigma}_{k}$ and $\bar{\sigma}^{*}_{k}$ with
the initial $\initi$ and final $\mid\Psi_{f}>$  states. In order to
obtain from it a
self-consistent stationary phase approximation to time evolution of
a given initial state $(\initi = \inito \otimes \initt)$, we
choose, as in reff\cite{cp}, the particular final state,

\beq
\mid\Psi_{f}> = \hat{\cal{U}}^{(0)}(t,0) \initi,
\eeq
where the zeroth-order evolution operator $\hat{\cal{U}}^{(0)}(t,0)$ is

\beq
\hat{\cal{U}}^{(0)}(t,0) = \tau \prod^{N}_{j=1}(1- i\epsilon \hat{h}_{j}), \\
\eeq
$\hat{h}_{j}$, is the total hamiltonian of the
two subsytem, eq(12). Substituing (13), in (9) and (10)
the following expression are found to the self-consistent
stationary phases approximation:

\beqa
\bar{\sigma}^{*}_{k} & =      &\limite
2 <\Psi_{i}\mid \hat{\cal{U}}^{(0)}(0,t_{k})\sumacc_{k}
\hat{\cal{U}}^{(0)}(t_{k},0) \initi \nonumber \\
                     & \equiv & 2u^{*}<\hat{\eta}^{\dagger}>_{k}+
2 v^{*}<\hat{\rho}^{\dagger}>_{k}
\eeqa

and

\beqa
\bar{\sigma}_{k} &  =     &\limite  2 <\Psi_{i}
\mid\hat{\cal{U}}^{(0)}(0,t_{k})
\suma_{k} \hat{\cal{U}}^{(0)}(t_{k}, 0) \initi \nonumber \\
                 & \equiv & 2 u<\hat{\eta}>_{k} +
2 v<\hat{\rho}>_{k}
\eeqa
where we have assumed that $\hat{\cal{U}}^{(0)}(t,0)$ is a unitary operator.
This, of course, depend of the particular choice  made to the auxiliary
matrix, $ u $, $v$. The expression to $\sigma_{k}$ and $\sigma^{*}_{k}$,
is given by the sum of the expection values of operators $(1)$ and $(2)$.
We can
also observe the separation between the operator $(1)$ and $(2)$,
(identifing $(1)$ with operators of system $1$ and $(2)$ with
operators of system $2$) in SPA.

 Assuming the  unitarity and replacing the expression to $\sigma_{k}$
and $\sigma^{*}_{k}$  in (12-a) and (12-b),
the mean-field  hamiltonian for the both subsystem  are the following:
\beqa
\hat{h}_{{1}_{j}} & = & \hat{H}^{1}_{1} + \nonumber \\
                  &   & +2 u^{*}<\hat{\eta}^{\dagger}>^{0}v
\hat{\rho} +
2 u<\hat{\eta}>^{0}v^{*}\hat{\rho}^{\dagger}
\;\;\;\; (a) \nonumber \\
                  &   & +2 v^{*}<\hat{\rho}^{\dagger}>^{0}v
\hat{\rho} +
2 v<\hat{\rho}>^{0}v^{*}\hat{\rho}^{\dagger}
 \;\;\;\; (b)\nonumber \\
\hat{h}_{{2}_{j}} & = & \hat{H}^{1}_{2} +  \nonumber \\
                  &   & +2 v^{*}<\hat{\rho}^{\dagger}>^{0}u
\hat{\eta} +
2 v<\hat{\rho}>^{0}u^{*}\hat{\eta}^{\dagger}
 \;\;\;\;  (c) \nonumber \\
                  &   & + 2 u^{*}<\hat{\eta}^{\dagger}>^{0}
u\hat{\eta}
+2 u<\hat{\eta}>^{0}u^{*}\hat{\eta}^{\dagger} \;\;\;\; (d)
\eeqa
or identifing the matrix products $u^{\dagger}v^{*}$, $v^{\dagger}v$,
$u^{\dagger}u$, $v^{\dagger}u$, etc with the two-body
"trial interaction"\cite{cp}
\beqa
\hat{h}_{{1}_{j}} & = & \hat{H}^{1}_{1} +\nonumber \\
                  &   & W_{12}<\hat{\eta}>^{0}\hat{\rho} +
W^{*}_{12}<\hat{\eta}^{\dagger}>^{0}\hat{\rho} +
\;\;\;\;\;\;\;\;\;\;\;\;\;\;\;\;  (a) \nonumber \\
                  &   & W_{1}<\hat{\rho}>^{0}
\hat{\rho}^{\dagger} +
W_{1}^{*}<\hat{\rho}^{\dagger}>\hat{\rho}  \nonumber\\
                  &   &    \nonumber\\
\hat{h}_{{2}_{j}} & = & \hat{H}^{1}_{2} + \nonumber \\
                  &   & W_{21}<\hat{\rho}>^{0}\hat{\eta}^{\dagger} +
W^{*}_{21}<\rho^{\dagger}>^{0}\hat{\eta}+
\;\;\;\;\;\;\;\;\;\;\;\;\;\;\;\;  (b) \nonumber\\
                  &   &
W_{2}<\hat{\eta}>\hat{\eta}^{\dagger} +
W_{2}^{*}<\hat{\eta}^{\dagger}>\hat{\eta}
\nonumber \\
\eeqa
with,
\bc
\[
W_{{12}_{\gamma\delta, \gamma'\delta'}}  = 2 \sum_{\alpha\beta}
v^{\dagger}_{\gamma\delta,\beta\alpha}u_{\alpha\beta,\gamma'\delta'}
 \;\;\;\;\; (a) \]
\[ W_{{1}_{\delta\gamma'\gamma\delta'}} =  2 \sum_{\alpha\beta}
v^{\dagger}_{\gamma\delta,\beta\alpha}v_{\alpha\beta,\gamma'\delta'}
\;\;\;\;\; (b) \]
\[ W_{{2}_{\delta\gamma'\gamma\delta'}}
=  2 \sum_{\alpha\beta}
u^{\dagger}_{\gamma\delta,\beta\alpha}u_{\alpha\beta,\gamma'\delta'}
\;\;\;\;\; (c) \]
\[ {\;\;\;\;\;\;\;\;\;\;\;} W_{21} =  W^{\dagger}_{12}
\;\;\;\;\;\;\;\;\;\;\;\;\;\;\;\;\;\;\;\;\; (d)\;\;, \;\;\;\;\;\;\;\;\;\;
\;\;\;\; etc., \]
\ec

The following relations to $v$, and $u$ have being used:
\bc
\[ v_{\alpha\beta\gamma\delta} \;\; =  \;\;
v^{*}_{\beta\alpha\delta\gamma} \;\;\;\;\;  ; \;\;\;\;\;
u_{\alpha\beta\gamma\delta} \;\; = \;\;
u^{*}_{\alpha\beta\gamma\delta} \;\;\;\; (a) \]
\[ v^{T}_{\gamma\delta\alpha\beta} \;\; = \;\;
v_{\alpha\beta\gamma\delta}  \;\;\;\;\; ; \;\;\;\;\;
u^{T}_{\gamma\delta\alpha\beta} \;\; = \;\;
u_{\alpha\beta\gamma\delta} \;\;\;\; (b) \]
\ec

The expressions $(18)-(a)$ and$(18)-(b)$ are represented
the mean-field
hamiltonian in SPA, and contain simultaniously
the mean-field evolution of the subsytem $1$, $2$
and $1$-$2$. \\

The essential difference between the mean-field
approximation to one system case\cite{cp} and
two subsystem is the appearance of this
interaction between $1$ and $2$.
In the one system case we have only the mean-field
approximation of the one kind operators, p.e., the
terms in $W_{1}$ (in fermionic case) or the
terms $W_{2}$ but  does not appear the terms $W_{12}$
which is responsable for the mixing interaction between $1$ and $2$
subsystem.
However, in two subsystem case, we have  for each subsytem, (eq $(18)-a$
for subsystem $1$ and eq. $(18)-b$, for subsytem $2$)
the terms $W_{12}$, which is representing the "effect"
from the mean-field approximation of the one subsystem
over the other,  plus the mean-field approximation of
the self subsystem, i.e. $W_{1}$ to subsystem $1$ and $W_{2}$ to
subsystem  $2$. \\ \\

The "trial interaction", $W_{1}$, $W_{2}$ and $W_{12}$
can be choosen to guarantees the unitarity of the stationary phase
approximation of the time evolution operator and consequently
$\hat{h}_{{1}_{j}}$ and $\hat{h}_{{2}_{j}}$ go to be
an unitary operators.
This choice, can also to
depend of the interaccion between 1 and 2 subsystem and  in general one
would like to reproduce the best interaction between 1 and 2. \\

The appropiate choice of the trial interaction, garantee the
unitarity of evolution operators in mean-field approximation
and $\initi$ has a unitary evolution throught
$\hat{\it{U}}^{(0)}(t,0)$. \\ \\

When the operators, $\hat{\eta}$ and $\hat{\eta}^{\dagger}$
are representing
bosons operator, we can replace this operator by a new operator
$\hat{B}$ ($\hat{B}^{\dagger}$) where:
\bc
\[ {\hat{B}} = \hat{\eta} - <\bar{b}> \]
and
\[ {\hat{B}}^{\dagger} = \hat{\eta}^{\dagger} - <\bar{b}>^{*} \]
\ec
where $<\bar{b}>$ is a $c-number$ . This choice, for
the bosonic operator, allows for the dynamical treatment of coherent
condensates in mean-field approximation.

Choosing $<\bar{b}>$ conveniently, such that $\hat{h}_{{2}_{k}}$
result diagonal, we arraive at the following expression to
$\hat{h}_{{2}_{k}}$
and $\hat{h}_{{1}_{k}}$ :
\beqa
\hat{h}_{{2}_{k}} & = & \widehat{\tilde{H}}^{1}_{2} \nonumber \\
\hat{h}_{{1}_{k}} & = & \hat{H}^{1}_{1} - \frac{K_{2}}{2}(
\widetilde{<\bar{b}>^{0}} + \widetilde{<\bar{b}>}^{{*}^{0}}) +
\nonumber\\
                  &   & W_{12}\widetilde{<\bar{b}>}^{0}\hat{\rho}^{\dagger}
+ W^{*}_{12}\widetilde{<\bar{b}>}^{0}\hat{\rho} + \nonumber \\
                  &   & W_{1}<\hat{\rho}>^{0} \hat{\rho}^{\dagger}
+ W^{*}_{1}<\hat{\rho}^{\dagger}>^{0}\hat{\rho} + \nonumber \\
\eeqa
with
\bc
\[ {\widehat{\tilde{H}}}^{1}_{2}  =  \sum_{p} K_{{2}_{p,p}}
\hat{B}^{\dagger}_{\bar{p}}\hat{B}_{p} \]
\ec
the diagonal hamiltonian of the subsystem $2$ and

\[ \widetilde{<\bar{b}>^{0}} \;\;\;\;\;\;\; ; \;\;\;\;\;\;\;
  \widetilde{<\bar{b}>^{{*}^{0}}} \]
the convenient choice to $<\bar{b}>$ such that $\hat{h}_{{2}_{k}}$
became diagonal. \\ \\

In the case of a fermionic system with two  different
interactions, $(V_{1}, V_{2})$, and  assuming $\hat{H}^{1}_{2}$,
$\hat{H}_{12}$ equal to zero, (consequently, $W_{12}$, is zero too).
The mean-field hamiltonian is given by:
\beq
{\hat{h}}_{{l}_{j}} = \hat{H}^{1}_{1} + W_{l}<\hat{\rho}_{l}>^{0}
\hat{\rho}^{\dagger}_{l} +  W^{*}_{l}<\hat{\rho}^{\dagger}_{l}>^{0}
\hat{\rho}_{l},
\eeq
with $l= 1,2$, the two different interactions and $\hat{\rho}_{l}$
replacing  $\hat{\rho}$ and $\hat{\eta}$. \\ \\

Finally, the general scheme used above  to obtain a mean-field approximation
to the time evolution of a given initial state can  be symple
extended to include higher order corrections to the SPA
The procedure, which was explain in detail in the reference\cite{cp}
(on one system case), can be summarized as follows. When the stationary
phase expansion of equations (9) and (10), is carried to order $n$,
we may identify an approximation of order $n$ to the evolution
operator, which we denote as $\hat{\cal{U}}^{(n)}(t,0)$,
as unitary operator. With
the help of this operator and impossing that the evolution of the
system  is given by $\hat{\cal{U}}^{(n)}(t,0)$, the self-consistent
choice which replaces equation (13) is
\beq
\mid \Psi_{f}> \rightarrow \hat{\cal{U}}^{(n)}(t,0) \mid \Psi_{i}>
\eeq
and allows on to define an n$th$ order corrected effective mean field
as:
\beq
\sigma^{(n)}_{k} = 2 \initii \hat{\cal{U}}^{(n)}(0,t_{k}) \suma_{k}
\hat{\cal{U}}^{(n)}(t_{k},0) \initi
\eeq
and an analogous expression to cc.,

\beq
\sigma^{{*}^{(n)}} = 2\initii \hat{\cal{U}}^{(n)}(0,t_{k})\sumacc_{k}
\hat{\cal{U}}^{(n)}(t_{k},0)\initi \\
\eeq

At the same time the initial state $\initi$ is the initial
state
$\initiio \initiit$ of the both subsystem and the
evolution operator $\hat{\cal{U}}^{(n)}(t,0)$, involves in general
an effective mean-field evolution governed by the total mean-field
hamiltonian,

\bc
\[ {\hat{h}}^{(n)}_{k} =
{\hat{h}}^{(n)}_{{1}_{k}} + {\hat{h}}^{(n)}_{{2}_{k}} \]
\ec
with
\beqa
{\hat{h}}^{(n)}_{{1}_{k}} & = & \hat{H}^{1}_{1}  +
\sigma^{(n)}_{k}v^{*}\hat{\rho}^{\dagger} +
\sigma^{{*}^{(n)}}_{k}v\hat{\rho} \;\;\;\;\;\;\;\;\;\; (a) \nonumber\\
{\hat{h}}^{(n)}_{{2}_{k}} & = & \hat{H}^{1}_{2} + \sigma^{(n)}_{k}
u^{*}\hat{\eta}^{\dagger} + \sigma^{{*}^{(n)}}_{k}u\hat{\eta}
\;\;\;\;\;\;\;\;\;\ (b) \nonumber\\
\eeqa

The nth order mean field approximation depend of the trial interaction
choosen and the evolution operator $\hat{\cal{U}}^{(n)}(t,0)$, will contain
appropriate correlation terms involving the operators of the $1$ and
$2$ subsystem of the interaction terms. \\

\section{Expansion of the evolution operator}

In this section we will extend the self-consistent initial condition
treatment of the preceding section to include higher corrections to
the simplest stationary phase approximation. As in ref.\cite{cp}
we will associate an approximation ${\it{U}}^{(n)}(t,0)$ to the
evolution operator corresponding to the estationary phase expansion
carried to order $n$
of the matrix element in eq. (6). This is done writing:

\beq
\final\hat{U}(t,0)\initi^{(n)} = \final\hat{\it{U}}^{(n)}(t,0) \initi
\eeq

This expansion involves, of course, a stationary phase path,
$\bar{\sigma}_{k}$ and $\bar{\sigma}^{*}_{k}$
which depends on both initial and final states and
of the subsystem $1$ and $2$. Requering further that the final state
$\mid \Psi_{f}>$ is that which results from $\mid \Psi_{i}>$ with the
approximate evolution operator, one is led to the self-consistent initial
condition problem. The conexion between, $\hat{\it{U}}^{(n)}(t,0)$, and
$\hat{\cal{U}}^{(n)}(t,0)$ will be shwon explicity in the next subsections.

\subsection{Zeroth order, $\hat{\bf{\it{U}}}^{(0)}(t,0)$}

The zero-order to evolution operator is obtained imposing in matrix element
expansion the condition:
\beq
\final \hat{U}(t,0) \initi^{(0)} = \final \hat{\it{U}}^{(0)}(t,0)
\initi
\eeq
The l.h.s. of this expression is given by:
\beqa
\final U(t,0) \initi^{(0)} & = & \limite exp \{i \frac{\epsilon}{2}
\sum_{k} \mid \bar{\sigma}_{k}\mid^{2} \} \final\tau \prod^{N}_{j=1}
(1-i\epsilon\hat{h}_{j}) \initi \nonumber \\
                           & = & \limite exp \{i \frac{\epsilon}{2}
\sum_{k} \mid\bar{\sigma}_{k}\mid^{2}\} \final\hat{\cal{U}}^{(0)}(t,0)
\initi
\nonumber \\
\eeqa
Comparing (26) and (27) we identify the evolution operator
in zeroth-order expansion as:
\beq
\it{\hat{U}}^{(0)} (t,0) = \limite exp \{ i\frac{\epsilon}{2}
\sum_{k} \mid \bar{\sigma}_{k}\mid^{2}\} \hat{\cal{U}}^{(0)}(t,0)
\eeq
with the stationary phase $\bar{\sigma}_{k}$ and $\bar{\sigma}^{*}_{k}$
being given by eqs, (9) and (10).
The expression (28) show
that, in general, the evolution operators
in zero order expansion, $\hat{\it{U}}(t,0)$, differs from
$\hat{\cal{U}}(t,0)$ by overall (phase dependent) phase factor.
In this order,with the self-consistent condition,
the final state is
\[ \mid \Psi_{f}> = \hat{\it{U}}^{(0)} (t,0) \initi. \]

The stationay phase eqs. (15) and (16), depends on
$\hat{\cal{U}}^{(0)}$
alone. The additional phase factor in eq. (28) guarantees in particular
that the $\hat{\it{U}}^{(0)}(t,0)$ evolution of a stationary self-consistent
mean-field states $\final$ is given by a
phase factor, containing the mean-field
temporal evolution of the subsystem $1$, $2$ and $1-2$, that is,
\beqa
\frac{1}{2}\mid \bar{\sigma}_{k} \mid^{2} &  = & W_{12}<\hat{\eta}>^{0}
<\hat{\rho}^{\dagger}>^{0}
+ W^{\dagger}_{12}<\hat{\rho}>^{0}<\hat{\eta}^{\dagger}>^{0} \nonumber\\
                                          &    & W_{1}<\hat{\rho}>^{0}
<\hat{\rho}^{\dagger}>^{0} +  W_{2}<\hat{\eta}>^{0}
<\hat{\eta}^{\dagger}>^{0} \nonumber\\
\eeqa

In  special case mentioned before of two subsystem
composed by fermions and bosons where  bosons operators
were replaced by $\hat{B} + <\bar{b}> $, the stationary
self-consistent  phase is given by:

\beqa
\frac{1}{2}\mid \bar{\sigma}_{k} \mid^{2} &  = & W_{12}
\widetilde{<\hat{b}>^{0}} <\hat{\rho}^{\dagger}>^{0}+
W^{\dagger}_{12}<\hat{\rho}>^{0} \widetilde{<\hat{b}>^{{*}^{0}}}
\nonumber\\
                                          &    & W_{1}<\hat{\rho}>^{0}
<\hat{\rho}^{\dagger}>^{0} +  W_{2}\widetilde{<\bar{b}>^{0}}
\widetilde{<\bar{b}>^{{*}^{0}}} \nonumber\\
\eeqa

Apparently this expression does not depend explicitly of the
bosonic operators, but it is represented by $W_{12}$ interaction
as we should see later with  a specific example.  \\ \\

In the other special case, of N-fermions operators with  two differents
interactions, the expression (29) will be modified by the term
$W_{12}$,  which is null and the self-consistent phase factor
contain the mean-field temporal evolution of the subsystem, $(1)$ and
$(2)$,
\beq
\frac{1}{2}\mid \bar{\sigma}_{k}\mid^{2} = \sum_{l= 1,2}
W_{l} <\hat{\rho}_{l}>^{(0)}
<\hat{\rho}^{\dagger}_{l}>^{(0)}
\eeq

This expression differe of one system case by the apperance of the
temporal mean-field approximation  for each one of the  subsystem
considered in the initial system or included in lineal terms.

\subsection{Second Order, $ \hat{\it{U}}^{(2)}(t,0)$ }

The second order to evolution operators is obtained making $n=2$ in
the general expression to the matrix element expansion in SPA
and envolve second derivatives of $S$ with respect to the
$\xi $ and $\xi^{*}$, (see appendix A).

This second  order is a lowest correction to the
SPA and modify $\hat{\it{U}}^{(0)}(t,0)$ by a factor,
$[det S^{(2)}]$, explicitly :
\beqa
\final\hat{U}(t,0) \initi^{(2)} & = & \limite exp \{i S^{(0)} \}
det\left[ S^{(2)} \right]^{-1} \nonumber \\
                                & = & \limite exp \{i S^{(0)} \}
exp \{ - tr ln S^{(2)} \} \nonumber \\
\eeqa
with $S^{(0)}$ representing the action in zero order,
and  $S^{(2)}$ the action in  second order, given by,
\bc
\[ S^{(2)} = \delta_{{k_{1}},{k_{2}}} + i\epsilon
\overline{\overline{M}}(k_{1},k_{2}) -
i\epsilon (1- \delta_{{k_{1}},{k_{2}}})
\overline{\overline{D}}(k_{1},k_{2}) \]
\ec
where $\overline{\overline{M}}(k_{1},k_{2})$ is the matrix element
diagonal in the time and  can be expressed in a matrix representation as:
\beq
<\hat{H}_{{int}_{k_{1},k_{2}}}>
\left( \begin{array}{cc} \frac{1}{q} &  0 \\
                           0         & \frac{1}{q} \end{array}\right)
- \left( \begin{array}{c} \Lambda_{{k}_{1}} \\
\Lambda^{\dagger}_{{k}_{1}}  \end{array} \right)
\left(\begin{array}{ll} \Lambda^{\dagger}_{{k}_{2}} & \Lambda_{{k}_{2}}
\end{array} \right )
\eeq
where $\Lambda_{k}$ is a short notation to matrix elements:
\[ \final \hat{\cal{U}}(t_{f},t_{k}) \suma_{k} \hat{\cal{U}}(t_{k},0)
\initi \]
and \[ <\hat{H}_{int}> = \final \hat{\cal{U}}(t_{f},t_{k})
\hat{H}_{int} \hat{\cal{U}}(t_{k},0) \initi . \]

The expression (33) envolve the  terms
${\hat{H}_{int}} = \hat{H}^{2}_{1} +\hat{H}^{2}_{2} +\hat{H}_{12}$
that is, all possible terms
contained in the original two-body interaction, into  subsystem $1$,
$2$ and  between the subsystem $1$ and $2$
and also the product of lineal terms included to "hand" in the
funtional
integral and responsible for the "trial interaction".
All those terms can be arrange in three differents expression
of the follwing way:

One envolving only  the interaction between $1$ and $2$ operators
that is, $K_{12}$ and $W_{12}$, explicitly:

\beqa
&  \final \hat{\cal{U}}(t_{f},t_{k})
\left (\frac{K_{12}}{q}\hat{\eta}_{k}\hat{\rho}^{\dagger}_{k} +
\frac{K^{*}_{12}}{q}\hat{\rho}_{k}\hat{\eta}_{k}^{\dagger} \right )
 \hat{\cal{U}}(t_{k},0)
\initi - &  \nonumber \\
& \left( \frac{W_{12}}{2} \final \hat{\cal{U}}(t_{f},t_{k})\hat{\eta}_{k}
\hat{\cal{U}}(t_{k},0) \initi \final
\hat{\cal{U}}(t_{f},t_{k})
\hat{\rho}^{\dagger}_{k} \hat{\cal{U}}(t_{k},0)\initi +
\right.  &  \nonumber \\
&  \frac{W_{21}}{2} \final \hat{\cal{U}}(t_{f},t_{k})\hat{\rho}_{k}
\hat{\cal{U}}(t_{k},0) \initi \final
\hat{\cal{U}}(t_{f},t_{k})
\hat{\eta}^{\dagger}_{k} \hat{\cal{U}}(t_{k},0)\initi +
         & \nonumber \\
& \frac{W^{*}_{12}}{2} \final \hat{\cal{U}}(t_{f},t_{k})
\hat{\eta}^{\dagger}_{k} \hat{\cal{U}}(t_{k},0) \initi \final
\hat{\cal{U}}(t_{f},t_{k}) \hat{\rho}\hat{\cal{U}}(t_{k},0)
\initi + & \nonumber \\
& \left. \frac{W^{*}_{21}}{2} \final \hat{\cal{U}}(t_{f},t_{k})
\hat{\rho}^{\dagger}_{k} \hat{\cal{U}}(t_{k},0) \initi \final
\hat{\cal{U}}(t_{f},t_{k})
\hat{\eta} \hat{\cal{U}}(t_{k},0)\initi \right)
         & \nonumber \\
\eeqa
the  other terms  envolve
the two-body interaction from the subsystem $1$, $(V_{1})$,
and $2$, $(V_{2})$,  with
the corespondent trial interaction, $W_{1}$
$W_{2}$:
\beqa
& \frac{V_{1}}{q} \final \hat{\cal{U}}(t_{f},t_{k})
\hat{\Omega}_{1} \hat{\cal{U}}(t_{k},0)
\initi - &  \nonumber \\
& \left( \frac{W_{1}}{2} \final \hat{\cal{U}}(t_{f},t_{k})\hat{\rho}_{k}
\hat{\cal{U}}(t_{k},0) \initi \final
\hat{\cal{U}}(t_{f},t_{k})
\hat{\rho}^{\dagger}_{k} \hat{\cal{U}}(t_{k},0)\initi + \right. &
\nonumber \\
& \left. \frac{W^{*}_{1}}{2} \final \hat{\cal{U}}(t_{f},t_{k})
\hat{\rho}^{\dagger}_{k}
\hat{\cal{U}}(t_{k},0) \initi \final
\hat{\cal{U}}(t_{f},t_{k})
\hat{\rho}_{k} \hat{\cal{U}}(t_{k},0)\initi \right) & \nonumber \\
\eeqa
and
\beqa
& \frac{V_{2}}{q} \final \hat{\cal{U}}(t_{f},t_{k})
\hat{\Omega}_{2} \hat{\cal{U}}(t_{k},0)
\initi - &  \nonumber \\
& \left( \frac{W_{2}}{2} \final \hat{\cal{U}}(t_{f},t_{k})\hat{\eta}_{k}
\hat{\cal{U}}(t_{k},0) \initi \final
\hat{\cal{U}}(t_{f},t_{k})
\hat{\eta}^{\dagger}_{k} \hat{\cal{U}}(t_{k},0)\initi + \right. &
\nonumber \\
& \left. \frac{W^{*}_{2}}{2} \final \hat{\cal{U}}(t_{f},t_{k})
\hat{\eta}^{\dagger}_{k}
\hat{\cal{U}}(t_{k},0) \initi \final
\hat{\cal{U}}(t_{f},t_{k})
\hat{\eta}_{k} \hat{\cal{U}}(t_{k},0)\initi \right) & \nonumber \\
\eeqa

Choosing  the trial interaction $W_{12}$ as
$W_{12} = \frac{K_{12}}{q}$,  imposing the self-consistent
condition on $ \mid\Psi_{f}>$ and assuming unitarity of
$\hat{U}^{(2)}(t,0)$, the terms in (34) from interaction between
subsystem $(1)$ and $(2)$ can be written as:
\beqa
& \frac{K_{12}}{q} \left( < \hat{\eta}_{k} \hat{\rho}^{\dagger}_{k}>
- \frac{1}{2} ( < \hat{\eta}_{k}><\hat{\rho}^{\dagger}_{k}> +
< \hat{\rho}^{\dagger}_{k}>< \hat{\eta}_{k}> ) \right)  \nonumber\\
&
\frac{K^{*}_{12}}{q} \left( < \hat{\rho}_{k}\hat{\eta}^{\dagger}_{k} >
- \frac{1}{2} (<\hat{\rho}_{k}><\hat{\eta}^{\dagger}_{k}> +
 < \hat{\eta}^{\dagger}_{k}><\hat{\rho}_{k}>) \right) \nonumber\\
\eeqa
which is  equal to zero.
This choice to $W_{12}$ verify the condition inmpossed to the
unitarity of the evolution operators and is the best choice
to represent the interaction between the subsystem
$1$ and $2$.  \\

Analogously, one appropriate choice to $W_{1}$ and $W_{2}$,
together with the selfconsistent unitarity condition
make null the  direct  contribution of two-body interaction
in subsystem
$(1)$ and $(2)$ as shown in the following expression:
\beqa
& \left ( \frac{V_{1}}{q}<\hat{\Omega}_{1}> -
 \frac{W_{1}}{2}<\hat{\rho}><\hat{\rho}^{\dagger}> -
\frac{W^{*}_{1}}{2}<\hat{\rho}^{\dagger}><\hat{\rho}>  \right )
\nonumber\\
& \left( \frac{V_{2}}{q}<\hat{\Omega}_{2}> -
  \frac{W_{2}}{2}<\hat{\eta}><\hat{\eta}^{\dagger}> -
\frac{W_{2}^{*}}{2}<\hat{\eta}^{\dagger}><\hat{\eta}> \right)
\nonumber\\
\eeqa

This expression is analogous to the one system case\cite{cp}, where
we verify that with the appropriate choice to $W_{1}$ or $W_{2}$,
the contribution
in second order became null. \\

 Any other choice to $W_{12}$,
$W_{1}$ and $W_{12}$ correct the SPA mean fied as defined by the
trial interaction alone, as we can observe from zero order in (28).
\\ \\

The nondiagonal termporal terms,
$\overline{\overline{D}}(k_{1},k_{2})$,
 can also to be written into a matrix representation as:
\beq
\left( \begin{array}{cc} D(\Lambda_{{k}_{1}},
\Lambda^{\dagger}_{{k}_{2}}) &
D(\Lambda_{{k}_{1}}, \Lambda_{{k}_{2}}) \\
D(\Lambda^{\dagger}_{{k}_{1}},
\Lambda^{\dagger}_{{k}_{2}}) &
D(\Lambda^{\dagger}_{{k}_{1}}, \Lambda_{{k}_{2}}) \end{array} \right)
\eeq
with
\[D(\Lambda_{{k}_{1}},\Lambda^{\dagger}_{{k}_{2}}) \] a short notation to:
\beqa
& \final \ufk\suma_{{k}_{1}} \ukkk \sumacc_{{k}_{2}}
\ukki \initi  - \nonumber \\
& \left( \final \ufk \suma_{{k}_{1}}  \uki \initi \right. \nonumber\\
& \left. \final \ufkk\sumacc_{{k}_{2}} \ukki
\initi \right),  \nonumber \\
\eeqa
the same for the other terms. \\

Here, we can observe,
the appearance  of terms which envolve the trial interaction
$W_{12}$, $W_{1}$ and $W_{2}$,
and as on the diagonal temporal matrix, we go to analise each one
separately. Impossing the self-consistent condition on
$\mid \Psi_{f}>$ and assuming the unitarity of $\hat{U}^{(2)}(t,0)$
for each terms one it,
the contribution  of terms with $u$,$v^{*}$, ($W_{12}$), is:
\[
u \left( <\hat{\eta}_{{k}_{1}} \hat{\rho}^{\dagger}_{{k}_{2}}>  -
 <\hat{\eta}_{{k}_{1}}><\hat{\rho}^{\dagger}_{{k}_{2}}> \right) v^{*} +
 \]
\[ v \left( <\hat{\rho}_{{k}_{1}} \hat{\eta}^{\dagger}_{{k}_{2}}>
- <\hat{\rho}_{{k}_{1}}>
 < \hat{\eta}^{\dagger}_{{k}_{2}}> \right) u^{*}. \]

This expression  can be  zero or not
depending of the operator $(1)$ and $(2)$, the initial state
$\initi$
and the trial interaction $W_{12}$. For example, this expression
became zero for the case mentionated in the last section
when the operator $\hat{\eta}$ is a bosonic operators and
we do the replacement $\hat{B} = \hat{\eta} - <\bar{b}>$.
In this situation and supossing
the inital state a non correlated
state, the expression envolving  $W_{12}$ is zero
because there are not correlation between fermionics and
bosonics operators. \\

The other terms envolve the operators of subsystem
$(1)$ and $(2)$ with
trial interaction $W_{1}$ and $W_{2}$ respectively,
of the following way:
\[
v \left( <\hat{\rho}_{{k}_{1}}\hat{\rho}^{\dagger}_{{k}_{2}}> -
<\hat{\rho}_{{k}_{1}}><\hat{\rho}^{\dagger}_{{k}_{2}}> \right)v^{*}
\]
and a similar expression to the other operators:
\[
u\left( <\hat{\eta}_{{k}_{1}}\hat{\eta}^{\dagger}_{{k}_{2}}> -
<\hat{\eta}_{{k}_{1}}><\hat{\eta}^{\dagger}_{{k}_{2}}> \right)u^{*}
\]
giving a RPA for each subsytems. The one system case is  a
particular case of this expression and  in general, it is a no null
expression because the correlations of the self operators $(1)$
and $(2)$. \\ \\

In the particular fermionic case with two differents interactions
$(V_{1}, V_{2})$,  the expression (37) is zero because
$W_{12}$ is zero and (38) is null as a consequence trial interaction
choice.
The essential difference with other special case  come
from   $\overline{\overline{D}}(k_{1},k_{2})$ contribution,
where this term would be  responsible for a RPA
 involving
$W_{1}$, $W_{2}$ and mixing terms with $W_{1}$ and $W_{2}$.
\\ \\

The expression in second order eq.(32) reduces to:
\beqa
\final\hat{U}(t,0) \initi^{(2)} & = & \limite \left (
exp \{i \frac{\epsilon}{2}
\sum_{k} \mid\bar{\sigma}_{k}\mid^{2} \} \final\hat{\cal{U}}^{(0)}(t,0)
\initi \right. \nonumber\\
                                &   & \left.
 exp \{ tr ln [ 1 + i\epsilon
(1 - \delta_{k_{1}k_{2}}) \overline{\overline{D}}(k_{1},k_{2})] \}
\right ).
\eeqa

 The operators, $\hat{\it{U}}^{(2)}(t,0) $, can  be identify as:
\beq
\hat{\it{U}}^{(2)}(t,0) = exp \{ \sum^{\infty}_{n=2} \frac{1}{n}
tr\left( i\epsilon \overline{\overline{D}}(k_{1},k_{2}) \right)^{n} \}
\hat{\it{U}}^{(0)}(t,0)
\eeq
where we have expanded
$ ln [ 1 + i\epsilon
(1 - \delta_{k_{1}k_{2}}) \overline{\overline{D}}(k_{1},k_{2})] $
and have started  the sum by $n=2$ in the expansion because the first
terms is zero in view of the trace in the time labels.

The corresponding self-consistency condition on
$ \mid \Psi_{f} > = \mid \Psi^{1}_{f} > \mid \Psi^{2}_{f} > $
now reads
\beq
\mid \Psi_{f} > = \hat{\it{U}}^{(2)}(t,0) \initi.
\eeq

The difference between the second order evolution operator
$\hat{\it{U}}^{(2)}(t,0)$ and
$\hat{\cal{U}}^{(0)}(t,0)$ consists therefore in two overall
phase-factor. The stationary path in second order, $\sigma^{(2)}$ and
$\sigma^{{*}^{(2)}}$, thus coincide with zero order as given by
equations $(17)$ and $(18)$, i.e. the mean field in this order
depends on $\hat{\cal{U}}^{(0)}(t,0)$ alone.

The content of eq (42) can be made explicit by  defining an
operator (see Appendix A)
\beqa
\overline{\hat{\bf{\Gamma}}}(t_{1},0)
& = & \hat{\rho}_{j}(t_{1}) \left \{ W_{j} \initi \initii \right.
\nonumber\\
&   & + \frac{i}{2} W_{j} \int^{t_{1}}_{0} dt_{2} \hat{\rho}_{l}(t_{2})
\initi \initii \hat{\rho}_{l}(t_{2}) W_{l} \nonumber\\
&   &  + \frac{i}{3} W_{j} \int^{t_{1}}_{0} dt_{2}
\hat{\rho}_{l}(t_{2})
\initi \initii \hat{\rho}_{l}(t_{2}) W_{l} \nonumber\\
&   &
i \int^{t_{2}}_{0} dt_{3} \hat{\rho}_{m}(t_{3})
\initi \initii \hat{\rho}_{m}(t_{2}) W_{m} \nonumber\\
&   & \left. + ....... \right \} \hat{\rho}_{j}(t_{1}),
\eeqa
 where $j,l,m = 1,2$  indices representing the
two differents
interaction $W_{1}$ and $W_{2}$, and
in terms of which (43) becomes
\[ \mid\Psi_{f}> = exp \left \{ i \int^{t}_{0} dt_{1} <\Psi_{f} \mid
\overline{\hat{\bf{\Gamma}}}(t_{1},0) \initi \right\}
\hat{\cal{U}}^{(0)}(t,0) \initi.  \]

This show that the self-consistent choice for $ \final$ to second order
involves a time evolution give by the self-consistent mean-field
$\hat{\cal{U}}^{(0)}(t,0)$ modify by the overall phase involving
loops fluctuation.

\subsection{Fourth order, $\hat{\it{U}}^{(4)}(t,0)$ }

This order involve fourth derivatives of $S$ with respect to $\xi_{k}$ and
$\xi^{*}_{k}$, in an expansion of the exponential factor in (7),
as show in appendix A,
\beq
\final \hat{U}(t,0) \initi^{(4)}  = \limite \final
\hat{\it{U}}^{(0)}(t,0) \initi \left( 1 + iS^{(4)} \right)
\eeq

where $S^{(4)}$ involves $\delta^{4}S[\xi,\xi^{*}]/\delta\xi_{{k}_{1}}
\delta\xi_{{k}_{2}}\delta\xi^{*}_{{k}_{3}}\delta^{*}_{{k}_{4}}$. \\

As mentioned before, the fourth order approximation differ from
zero  and second order in that it no longer consists of a phase-factor
which appear modifying the mean-field evolution operators
$\hat{\it{U}}^{(0)}(t,0)$. Realy, in this order appear the
correlation terms form the system $1$ and $2$, in the evolution
operators in the  form
\beq
\hat{\it{U}}^{(4)} = \hat{\cal{U}}^{(0)}_{4}(t,0) +
\hat{\cal{U}}^{(4)}(t,0).
\eeq
$\hat{\it{U}}^{(4)}(t,0)$ is a unitary operators. The operator
$\hat{\cal{U}}^{(0)}_{4}(t,0) = \prod^{N}_{k=1}(1- i\epsilon\hat{h}_{k})$
is a mean-field operator in fourth order,  but differs
from the zero order, $\hat{\cal{U}}^{(0)}(t,0)$, as it contains
information on the correlation between subsytem $1$ and $2$ through
$\hat{h}_{k}$ which now contains the fourth order approximation
to the auxiliary field $\sigma$ and $\sigma^{*}$, i.e.
\[
{\hat{h}}_{k}  =  \hat{h}_{{k}_{1}} + \hat{h}_{{k}_{2}} \]
with.

\beqa
\hat{h}_{{k}_{1}} & = & \hat{H}^{1}_{1} +
\sigma^{{*}^{4}}_{k}v \hat{\rho}
+\sigma^{4}_{k}v^{*} \hat{\rho}^{\dagger} \;\;\;\;\;\; (a)\nonumber \\
\hat{h}_{{k}_{2}} & = & \hat{H}^{1}_{2} +
\sigma^{{*}^{4}}_{k}u \hat{\eta}
+\sigma^{4}_{k}u^{*} \hat{\eta}^{\dagger} \;\;\;\;\;\; (b) \nonumber\\
\eeqa
for the special case of an uncorrelated state $ \initi$,
$\hat{h}_{{k}_{1}}$ and $\hat{h}_{{k}_{2}}$ can be written as:
\beqa
\hat{h}_{{1}_{k}} & = & \hat{h}^{(0)}_{{1}_{k}} +
\hat{h}^{(4)}_{{1}_{k}}  \;\;\;\;\; (a)\nonumber\\
\hat{h}_{{2}_{k}} & = &
\hat{h}^{(0)}_{{2}_{k}}+ \hat{h}^{(4)}_{{2}_{k}} \;\;\;\;\; (b)
\eeqa
where $\hat{h}^{(0)}_{{i}_{k}}$ is the uncorrelated part of
the mean field hamiltonian ,  analogous at the
$\hat{h}_{{i}_{k}}$ defined in equation (18) which appear in the mean
field evolution operator $\hat{\cal{U}}^{(0)}(t,0)$. The second terms,
$\hat{h}^{(4)}_{{i}_{k}}$ can be written as:
\beqa
\hat{h}^{(4)}_{{1}_{k}} & = & \left( W_{12} <\Psi_{i} \mid
\hat{\cal{U}}^{(0)}_{4}(0,t_{k}) \hat{\eta}_{k}
\hat{\cal{U}}^{(4)}(t_{k},0) \mid \Psi_{i}> \hat{\rho}^{\dagger}
\right.  \nonumber\\
                        &   & \left. +  W^{*}_{12} <\Psi_{i} \mid
\hat{\cal{U}}^{(0)}_{4}(0,t_{k})\hat{\eta}^{\dagger}_{k}
\hat{\cal{U}}^{(4)}(t_{k},0) \mid \Psi_{i}> \hat{\rho}
\right)   \nonumber\\
                        &   & + \left( W_{1} <\Psi_{i} \mid
\hat{\cal{U}}^{(0)}_{4}(0,t_{k}) \hat{\rho}_{k}
\hat{\cal{U}}^{(4)}(t_{k},0) \mid \Psi_{i}> \hat{\rho}^{\dagger}
\right.  \nonumber\\
                        &   & \left. + W_{1}^{*}<\Psi_{i} \mid
\hat{\cal{U}}^{(0)}_{4}(0,t_{k}) \hat{\rho}^{\dagger}_{k}
\hat{\cal{U}}^{(4)}(t_{k},0) \mid \Psi_{i}> \hat{\rho}
\right) \\
                        &   & \nonumber\\
\hat{h}^{(4)}_{{2}_{k}} & = & \left( W_{21} <\Psi_{i} \mid
\hat{\cal{U}}^{(0)}_{4}(0,t_{k}) \hat{\rho}_{k}
\hat{\cal{U}}^{(4)}(t_{k},0) \mid \Psi_{i}> \hat{\eta}^{\dagger}
\right.  \nonumber\\
                        &   & \left. + W^{*}_{21} <\Psi_{i} \mid
\hat{\cal{U}}^{(0)}_{4}(0,t_{k})\hat{\rho}^{\dagger}_{k}
\hat{\cal{U}}^{(4)}(t_{k},0) \mid \Psi_{i}> \hat{\eta}
\right)   \nonumber\\
                        &   &  \nonumber \\
                        &   & + \left( W_{2} <\Psi_{i} \mid
\hat{\cal{U}}^{(0)}_{4}(0,t_{k}) \hat{\eta}_{k}
\hat{\cal{U}}^{(4)}(t_{k},0) \mid \Psi_{i}> \hat{\eta}^{\dagger}
\right.  \nonumber\\
                        &   & \left. + W_{2}^{*}<\Psi_{i} \mid
\hat{\cal{U}}^{(0)}_{4}(0,t_{k}) \hat{\eta}^{\dagger}_{k}
\hat{\cal{U}}^{(4)}(t_{k},0) \mid \Psi_{i}> \hat{\eta}
\right )
\eeqa

The evolution operator $\hat{\cal{U}}^{(4)}(t,0)$, is the responsable
for two-body $(1),(2)$
correlations and  between 1-2 operators correlations begin  given by
\beq
\hat{\cal{U}}^{(4)}(t,0) = \hat{\cal{U}}^{(0)}_{4}(t,0) \frac{i}{6}
\int^{t}_{0}dt_{2}\int^{t_{1}}_{0}dt_{1} \frac{3}{4}
\left( \hat{H}_{int}(t_{2}) \hat{H}_{int}(t_{1}) \right)_{linked},
\eeq
where, $\hat{H}_{int}(t)$, is a short notation for
\[ {\hat{\cal{U}}^{(0)}_{4}(0,t)}\hat{H}_{int}
\hat{\cal{U}}^{(0)}_{4}(0,t) \]
or in terms of the respectives two-body operators:
\[ {\hat{\cal{U}}^{(0)}_{4}(0,t)} (V_{1}\hat{\Omega}_{1}
 +V_{2}\hat{\Omega}_{2} +(K_{12}\hat{\rho}\hat{\eta}^{\dagger} +
K_{12}^{*}\hat{\eta}\hat{\rho}^{\dagger}) )
\hat{\cal{U}}^{(0)}_{4}(t,0) .\]

The self-consistent final state $\mid \Psi_{f}>$ is again constructed as
\[ {\mid\Psi_{f}>} = \hat{\it{U}}^{(4)}(t,0) \Psi_{i}>. \]

The effect of the second term in fourth order evolution correlation
operators, eq (45), will be to generate two-body correlation $(1), (2),
\;\;\;\; and \;\;\;\; (1-2) $
operators during the propagation of this initial uncorrelated state
as show in fig.2 \\ \\

We now go to analise  the fermionic (1) plus bososic (2) case,
using the new bosonic
operator $\hat{B} = \hat{\eta} - <\bar{b}> $. Again
$\hat{h}_{{2}_{k}}$ is diagonal and equal to $\hat{h}^{0}_{{2}_{k}}$
with $\hat{h}^{4}_{{2}{k}} $ zero. The terms $\hat{h}^{0}_{{1}_{k}}$
is equal to (19) and
\beqa
\hat{h}^{4}_{{1}_{k}} & = &
W_{12}\widetilde{<\bar{b}>^{(4)}}\hat{\rho}^{\dagger}
+ W^{*}_{12}\widetilde{<\bar{b}>^{{*}^{(4)}}}\hat{\rho} +
\nonumber \\
                      &   & W_{1}<\hat{\rho}>^{(4)}
\hat{\rho}^{\dagger}
+ W^{*}_{1}<\hat{\rho}^{\dagger}>^{(4)} \hat{\rho}  \nonumber \\
\eeqa
with
\[ <\bar{b}> \;\;\; = \;\;\; \widetilde{<\bar{b}>^{(0)}}  \;\;\;
 + \;\;\; \widetilde{<\bar{b}>^{(4)}} \]
the convenient choice to $<\bar{b}>$ such that $\hat{h}_{{2}_{k}}$
became diagonal. \\

For the fermionic system with two different interactions the
mean-field contribution in zero order, $\hat{h}^{(0)}_{{l}_{k}}$
is analogous at (20), furthermore
\[ {\hat{h}^{(4)}_{{l}_{k}}} = W_{l}<\hat{\rho}_{l}>^{(4)}>
\hat{\rho}^{{(4)}^{\dagger}}_{l} +
W^{*}_{l}<\hat{\rho}^{\dagger}_{l}>^{(4)}>
\hat{\rho}^{(4)}_{l} } \]
correlation of fourth order.

\section{A Mean-Field Approximation in $\sigma$-model: A Didatic
Example}

With the idea to illustrate and explain, of a didatic way,
the effects of  the mean-field
approximation and  the self-consistent expansion obtained in
the last section
we should apply this formalism (self-consistent MFA)
in a $\sigma$- model system.
This model represent a system composed by fermions and mesons
fields and we go to show how the self-consistent
MFA modify the fermions mass  as a consequence of
 mesons scalar field (bosons) presence. \\
As said in the beginning, the idea of this example is
explain in a didatic way the effect of the self-consistent
mean-field expansion for the evolution operator and understand
how are this effects and how appear the correlation in the temporal
evolution of the two (or more) subysytem with a interaction between
their.
The Lagrangian
dentsity for the $\sigma$ model is given by:
\[
{\it{L}}_{\sigma}  =  \bar{\psi}
[\gamma_{\mu}(i\partial^{\mu}  - (m-g_{s}\phi_{s})]\psi
+\frac{1}{2}
(\partial_{\mu}\phi_{s}\partial^{\mu}\phi_{s} - m_{s}^{2}\phi^{2}) \]

We consider an infinite system and use the following
expansion for the fermionic and mesonic fields:

\beqa
\psi         & = & \sum_{k,l,s} [\frac{(2\pi)^{3}}{V}
\frac{m}{E_{\circ}(\vec{k})}]^{1/2}u_{l}(\vec{k}, s)c_{l}(\vec{k}, s)
exp\{-i(-)^{l}k. x\}    \\
\phi_{s} & = & \sum_{p} [\frac{1}{\sqrt{2v}\alpha_{p}}]
[\hat{b}_{p} exp\{ ip.x \} + \hat{b}^{\dag}_{p}
exp\{ -ip.x \} ]
\eeqa

with $(l= 0,1)$ and
\[
u_{\circ}(\vec{k},s)  =  \sqrt{\frac{E_{\circ}(\vec{k})+m}{2m}} \left(
\begin{array}{c}
\chi_{s}  \\
\frac{\vec{\sigma}.\vec{k}}{E_{\circ}(\vec{k})+m} \chi_{s}
\end{array}\right),\]

\[ u_{1}(\vec{k},s)  =  \sqrt{\frac{E_{\circ}(\vec{k})+m}{2m}} \left(
\begin{array}{c}
\frac{- \vec{\sigma}.\vec{k}}{E_{\circ}(\vec{k})+m} \chi_{s} \\
\chi_{s}
\end{array}\right) \]
the Dirac Spinors of the positive and negative energies
respectively, with :
\[
{\bar{u_{l'}}}(\vec{k},s')u_{l}(\vec{k},s) =
\{(-)^{l}\delta_{ll'} - \frac{\vec{\sigma}.\vec{k}}{m}(1 - \delta_{ll'})\}
\delta_{ss'} \]

\[ u^{\dag}_{l'}(\vec{k},s')u_{l}(\vec{k},s)  =
\frac{E_{\circ}(\vec{k},s)}{m}\delta_{ll'}\delta_{ss'}. \]

 In this representation, the operator,  $ c_{0}(\vec{k}, s) $
annihilates a fermion with $ E_{\circ}(\vec{k}) >0  (particles) $,
$ c_{1}(\vec{k}, s) $ creates a fermion with $E_{\circ}(\vec{k}) <0
(antiparticles) $ and $ b $ is the boson operators and is represented
the $\sigma $ meson .
The $|vac>$ state for noninteracting fields is defined as:
$c_{\circ}(\vec{k},s)|vac> = c^{\dag}_{1}(\vec{k},s)|vac> = 0$. \\ \\

    The Hamiltonian is obtained through $ H_{\sigma}
 = \int d^{3}x \it{H}(x) $ where, after replace (53) and (54) in
$\it{H}$,
we can identify the  terms: $ H_{1} + H_{2} + H_{int}$ as :
\beqa
H_{1} = K_{1}\hat{\rho}
& ;
&   \nonumber\\
& \left(K_{1} \right)_{\alpha \beta}  =
& \{ \frac{(2\pi)^{3}}{V}
\frac{m\delta(\vec{k_{\beta}}-\vec{k}_{\alpha})}
{[E_{\circ}(\vec{k}_{\alpha})E_{\circ}(\vec{k}_{\beta})]^{1/2}}
\nonumber \\
&
& exp\{ -i[(-)^{l_{\beta}}k^{\circ}_{\beta}
-(-)^{l_{\alpha}}k^{\circ}_{\alpha}]x_{\circ} \nonumber \\
&
&  \bar{u}_{l_{\alpha}}(\vec{k}_{\alpha}, s)
(\vec{\gamma}.\vec{k}_{\beta} + m)
u_{l_{\beta}}(\vec{k}_{\beta},s') \}  \nonumber \\
&
& \nonumber\\
& \hat{\rho}_{\alpha\beta} =
& c^{\dag}_{l}(\vec{k},s)c_{l'}(\vec{k'},s') \nonumber \\
&
&    \nonumber \\
H_{2} = K_{2}\hat{b}^{\dagger}\hat{b}
& =
& \sum_{p} (p^{\circ})^{2} \hat{b}^{\dagger}(p)\hat{b}(p)
 \nonumber \\
& K_{2} =
&   (p^{\circ})^{2} \nonumber \\
&
& \nonumber\\
H_{int} = H_{12} = K_{{12}_{+}}\hat{\rho}^{\dagger}\hat{b} +
K_{{12}_{-}}\hat{\rho}\hat{b}^{\dagger}
& ;
& \nonumber \\
& \left( K_{12}\right)_{\pm(\gamma\delta)} =
& \{ -g_{\sigma} [ \frac{(2\pi)^{3}}{V} ]
\frac{m}{[E_{\circ}(\vec{k_{\gamma}})E_{\circ}(\vec{k}_{\delta})]^{1/2}}
\nonumber\\
&
& exp\{ -i[(-)^{l_{\delta}}k^{\circ}_{\delta}
- (-)^{l_{\gamma}}k^{\circ}_{\gamma} - (\pm) p^{\circ}]x_{\circ} \}
\nonumber \\
&
& \bar{u}_{l_{\gamma}}(\vec{k_{\gamma}}, s)
u_{l_{\delta}}(\vec{k_{\delta}}, s')
\delta\left( \pm (\vec{k}_{\delta} - \vec{k}_{\gamma} - \vec{p}) \right) \}
\nonumber\\
&
& \nonumber\\
&  \hat{\rho}^{\dagger}\hat{b} =
& c^{\dagger}_{l_{\gamma}}(\vec{k}_{\gamma}, s)
c_{l_{\delta}}(\vec{k}_{\delta}, s') \hat{b}(\vec{p})
\nonumber \\
\eeqa

\subsection{Self-Consistent  Mean Field Approximation}

The self-consistent MFA is obtained in a direct way through
(18) or (19) expression,
\beqa
\hat{h}_{{2}_{k}} & = & \widehat{\tilde{H}}_{2}
- \frac{K_{2}}{2}(
\widetilde{<\bar{b}>^{0}} + \widetilde{<\bar{b}>^{{*}^{0}}})
\nonumber \\
\hat{h}_{{1}_{k}} & = & \hat{H}_{1} + \nonumber \\
                  &   & W_{12}\widetilde{<\bar{b}>^{0}}
\hat{\rho}^{\dagger}
+ W^{*}_{12}\widetilde{<\bar{b}>^{{*}^{0}}}\hat{\rho} + \nonumber \\
                  &   & W_{1}<\hat{\rho}>^{0} \hat{\rho}^{\dagger}
+ W^{*}_{1}<\hat{\rho}^{\dagger}>^{0}\hat{\rho} \nonumber \\
\eeqa

Assuming the ground state, $ |\Psi_{i}> $, as a direct
product of the fermionic and mesonic ground states, i.e.,
$|\Psi_{i}> = |\Psi^{1}_{i}> \otimes |\Psi^{2}_{i}> $
 and in particular  \[ |\Psi^{1}_{i}> =
\prod_{{k,k'\leq k_{F}}} c^{\dagger}_{0}(\vec{k}, s)
c_{1}(\vec{k'}, s') |vac>, \]
 the expression to $\widetilde{<\bar{b}>^{0}}$, given by
\bc
\[
{\widetilde{<\bar{b}>^{0}}} =  - \frac{W^{\dagger}_{21}}{K_{2}
+ W_{2}}  <\hat{\rho}>  \]
\[
{\widetilde{<\bar{b}>^{{0}^{*}}}}  =  - \frac{W^{T}_{21}}{K_{2}
+ W_{2}^{*}} <\hat{\rho}^{\dagger}> \]
\ec
is:

\beq
\widetilde{<\bar{b}>^{0}} = - \frac{-g_{s}}{2}\sum_{\gamma'}
\frac{(2\pi)^{3}}{V} \frac{m_{N}}{E_{\circ}(\vec{k}_{\gamma'})}
\frac{1}{K_{{2}_{p}} + W_{2}} \Theta(k_{F}-\mid\vec{k}_{\gamma'}\mid)
\delta(\vec{p})exp\{ip^{\circ}.x_{\circ}\}
\eeq
and analogous expression to cc.

   As said earlier, the expression to
$<\bar{b}> $ depend essentially of fermionic ground state ,
as we can observe from  the result of (57), where also appear
the condition $ \delta(\vec{p})$ for the mesoninc quantum number.
Choosing, $W_{2} \;\; = \;\;0 $, identifing
$K_{{2}_{p=0}} $ with the scalar meson mass,  and writting
the sum as an integral, the expression (57) is given by:
\beq
\widetilde{<\bar{b}>^{0}} = \frac{g_{s}}{2}\frac{\rho_{s}(m)}{4 m_{s}}
\eeq
where we have identified the scalar density, as usually is done,  with :
\[ 4 \int d^{3}k \frac{m}{E^{\circ}(\vec{k})} \Theta (k_{F} - \vec{k})
\]
and analogous expression to cc.. \\

Obtained the expression to $\widetilde{<\bar{b}>^{0}}$ we  replace
its  in (56) and the mean-field hamiltonian to evolution
operator in mean-field approximation is given by:
\beqa
\hat{h}_{{2}_{k}} & = & \widehat{\tilde{H}}_{2}
- \frac{K_{2}}{2}(
\widetilde{<\bar{b}>^{0}} + \widetilde{<\bar{b}>^{{*}^{0}}}) +
  \nonumber \\
\hat{h}_{{1}_{k}} & = & \hat{H}_{1} + \nonumber \\
                  &   & - \frac{(2\pi)^{3}}{V} \sum_{k_{\gamma},k_{\delta}}
\frac{m \delta (\vec{k}_{\delta} - \vec{k}_{\gamma}}
{[E_{\circ}(k_{\gamma})E_{\circ}(k_{\delta})]^{1/2}}
exp\{ -i[(-)^{l_{\delta}}k^{\circ}_{\delta} -
(-)^{l_{\gamma}}k^{\circ}_{\gamma}]x_{\circ} \} \nonumber\\
                  &   &
( g^{2}_{s}\frac{\rho_{s}(m)}{4m_{s}} )
\bar{u}_{{l}_{\gamma}}(\vec{k}_{\gamma},s)
u_{{k}_{\delta}}(\vec{k}_{\delta},s')
c^{\dagger}_{{l}_{\gamma}}(\vec{k}_{\gamma},s)
c_{{l}_{\delta}}(\vec{k}_{\delta},s') \nonumber\\
\eeqa

As we can observe , the second terms in l.h.s. has the
same interaction
that the kinematic interaction, in $\hat{H}_{1}$, so
putting both expression together result:

\beqa
\hat{h}_{{1}_{k}} & = & \frac{(2\pi)^{3}}{V} \sum_{k_{\gamma},k_{\delta}}
\frac{m \delta (\vec{k}_{\delta} - \vec{k}_{\gamma}}
{[E_{\circ}(k_{\gamma})E_{\circ}(k_{\delta})]^{1/2}}
exp\{ -i[(-)^{l_{\delta}}k^{\circ}_{\delta} -
(-)^{l_{\gamma}}k^{\circ}_{\gamma}]x_{\circ} \nonumber\\
                  &   &
\bar{u}_{{l}_{\gamma}}(\vec{k}_{\gamma},s)
(\vec{\gamma}.\vec{k}_{\delta} + m^{*})
u_{{k}_{\delta}}(\vec{k}_{\delta},s')
c^{\dagger}_{{l}_{\gamma}}(\vec{k}_{\gamma},s)
c_{{l}_{\delta}}(\vec{k}_{\delta},s') \nonumber\\
\eeqa
where we are identifing the term, $m^{*}$, with:
\bc
\[ m^{*} =  m + g^{2}_{s}\frac{\rho_{s}(m)}{4m_{s}} \]
\ec
The selfconsistent mean-field condition to fermions and bosons,
and the appropriate  choise to $W_{12}$, make  possible to identify
the interaction between $(1)$ and $(2)$ subsystem in
mean-field hamiltonian
with  a modification in $H_{1}$ hamiltonian. This modification is
a consequency of mesons operators interacting with fermions operatos.
For this special conditions, i.e, the ground states fermionic
composited by particles and anti-particles with momentum $k$ below
to fermi sea and supossing no two-body bosonic interactions,  we
have reproduced the Walecka result, to effect mass, as a particular
case of general initial system. \\

Incluging of vector meson term in $H_{12}$ a vector correction for
fermionic interation is obtained of straightforward way\cite{cn}
self-consistent main field approximation.

\section{Conclution}

In this work was developed a self-consistent mean-field
expansion for a
general system composed by two subsystem  with initial condition
on the basis of the flexible functional integral representation
proposed by Kerman, Levit and Troudet\cite{klt} and used by ours
in recent papers
to develope a mean-field expansion for the many-fermions initial
condition problem\cite{cp}.

The system consider here  is composed by two general
subsystem, $(1)$ and $(2)$, containing  each subsytem
 a kinetic plus
two body interaction term and an additional interaction term between
their. This is a general system  and
the formalim makes possible to generalize it  for system with
more of two subsystem or interaction. This possible generalization
is the result of lineal term put to "hand" in funtional integral
 formulation (2), (3).

To each order in the expansion, one considers the transition
amplitude from the prescribed initial state
to the final state to which it self-consistently evolves. This
condition is essential to obtain a self-consistent mean-field
expansion. In zero order one obtain a time dependent mean- field
which have information about interaction of the total system, which
include the subsystem $(1)$, $(2)$ and the interaction term between
their. It is important to mention that in this order
the all two-body interaction term appear with trial interaction
which  is a consequencia of lineal term  put to "hand"
in funtional integral. In second order the evolution operator
contain the RPA correlation effects (or one  loops expansion)
in the factor phase.
This RPA envolve the different interaction term from subsystem
$(1)$ and $(2)$ and a mixing interaction terms as observed
in fig.1. The fourth order expansion contain explicitly
the "fourth" order correction through $\hat{\cal{U}}^{(4)}(t,0)$,
where two body correlation appear explicitly envolving
fermions and bosons operators  as shown in fig.2 .
The same correlations
effects also modify the effective
mean field  hamiltonian approximations in this order.

The effects of this self-consistent mean-field approximations
in lowest order
is shown in $\sigma$-model, where the scalar mesons are
treated as quantum particles, resulting an effective mass
for fermions (nucleons) as a consequencia of
interactions term between fermions (nucleons) and mesons (bosons)
as soon as by  the special intial state for both subsystem.

The general chacarteristics of the systems and expansion
make possible  to apply this results into  any system composed
by two or more subsystem such that mentioned before
and  also to any system
for which is possible to distinguish between two or
more subsystem, (p.e. collision term in composed
systems),  as soon as in composed system temperature
dependent.

The effect of self-consistent mean field hamiltonian in
fourth order for $\sigma$-model and NJL in all order is  being
elaborated
and as immediate step a description of the maser model with this
formalism would be an intersting and important topic to understand
the correlation effects in composed systems.

\section{Acknowledgements}

I would like to acknowledge the Depart. of Atomic Molecular and
Nuclear Physics group, from Sevilla University where this work was done.

\vspace{3cm}

\appendix{Appendix A}

The expansion of eq. (7) about stationary phase trajectoria is given
by:

\beq
\final \hat{U}(t,0)\initi = \limite e^{iS^{0}} N \int D[\xi, \xi^{*}]
exp \left \{ i\sum^{\infty}_{n=2}\left( \delta^{n} S \right)
(\bar{\sigma}, \bar{\sigma}^{*}) \right \}
\eeq

where the zero order terms is
\[ S^{0} = \frac{\epsilon}{2} \sum_{k} \mid \bar{\sigma} \mid^{2} -
i ln \final \hat{\cal{U}}^{(0)}(t,0) \initi \]
and
\[ {\hat{\cal{U}}}^{(0)}(t,0) = \tau \prod^{N}_{k=1}
\left( 1-i\epsilon (\hat{h}_{{1}_{k}} + \hat{h}_{{2}_{k}} ) \right) \]
with $\hat{h}_{{1}_{k}}$ and $\hat{h}_{{2}_{k}}$ given by  $18 -(a)$
and $18-(b)$ respectively. \\

The second order expression is obtained by keeping $n=2$  in the
exponential factor of (61), that is:

\beq
\final \hat{U}(t,0)\initi^{(2)} =
\limite e^{iS^{0}} N \int D[\xi, \xi^{*}] exp \left \{ \frac{i}{2}
\epsilon \sum_{k_{1},k_{2}} \Phi_{{k}_{1}} S^{(2)}(k_{1},k_{2})
\Phi^{\dagger}_{{k}_{2}} \right \}
\eeq
where $S^{(2)}(k_{1},k_{2})$ is a short notation to:\\

\[ {\left( \frac{\partial^{2}S[\sigma, \sigma^{*}]}
{\partial\xi_{{k}_{1}}\partial\xi_{{k}_{2}}}
\right )}_{\xi= \xi^{*}=0} \;\; , \;\;
\left( \frac{\partial^{2}S[\sigma, \sigma^{*}]}
{\partial\xi^{*}_{{k}_{1}}\partial\xi^{*}_{{k}_{2}}}
\right )_{\xi= \xi^{*} = 0} \;\; , \;\;
\left( \frac{\partial^{2}S[\sigma, \sigma^{*}]}
{\partial\xi_{{k}_{1}}\partial\xi^{*}_{{k}_{2}}}
\right )_{\xi= \xi^{*} = 0}  \]
and \[ {\Phi}_{k} = ( \xi_{k}, \xi^{*}_{k} ). \]

Resulting, after integration:
\beqa
\final\hat{U}(t,0) \initi^{(2)} & = & \limite exp \{i S^{(0)} \}
det\left[ S^{(2)} \right]^{-1} \nonumber \\
                                & = & \limite exp \{i S^{(0)} \}
exp \{ - tr ln S^{(2)} \} \nonumber \\
\eeqa
with the appropriated choice to trial interaction and
self-consistent condition  the evolution operator in
second order expansion  reduces  at
\beq
{\hat{\it{U}}}^{(2)}(t,0) = exp \{ \sum^{\infty}_{n=2} \frac{1}{n}
tr\left( i\epsilon(1 - \delta_{k_{1} k_{2}})
\overline{\overline{D}}(k_{1},k_{2}) \right)^{n} \}
\hat{\it{U}}^{(0)}(t,0)
\eeq
 The sum to $n=2$ in exponential factor can be express as:
\[ lim_{\epsilon \rightarrow \infty} \frac{1}{2}tr
\left \{(i\epsilon)^{2}(1- \delta_{k_{1}k_{2}})
\sum_{{k}_{2}} \overline{\overline{D}}(k_{1},k_{2})
(1- \delta_{k_{2}k_{3}})  \overline{\overline{D}}(k_{2},k_{3})
\right \}, \]
giving for the time trace
\[ lim_{\epsilon \rightarrow \infty} \frac{1}{2}Tr
\sum_{{{k}_{1},{k}_{2}}_{{k}_{1}\neq {k}_{2}}} \left \{(i\epsilon)^{2}
 \overline{\overline{D}}(k_{1},k_{2})
\overline{\overline{D}}(k_{2},k_{1}) \right \},  \]
expliciting one of the terms of the matrix elemnents for
$\overline{\overline{D}}(k_{1},k_{2})
\overline{\overline{D}}(k_{2},k_{1})$ we have
\[ <\Lambda_{{k}_{1}} \Lambda^{*}_{{k}_{2}}>
<\Lambda_{{k}_{2}}\Lambda^{*}_{{k}_{1}}> = \]
\[ <(v\hat{\rho}_{{k}_{1}}+ u\hat{\eta}_{{k}_{1}})
( v^{*}\hat{\rho}^{\dagger}_{{k}_{2}}+
u^{*}\hat{\eta}^{\dagger}_{{k}_{2}})>
<(v\hat{\rho}_{{k}_{2}}+ u\hat{\eta}_{{k}_{2}})
( v^{*}\hat{\rho}^{\dagger}_{{k}_{1}}+
u^{*}\hat{\eta}^{\dagger}_{{k}_{1}})> \]

or more explicitly:

\[ W_{1} <\hat{\rho}_{{k}_{1}} \hat{\rho}^{\dagger}_{{k}_{2}} >
<\hat{\rho}_{{k}_{2}} \hat{\rho}^{\dagger}_{{k}_{1}} >W_{1} + \]
\[ W_{2} <\hat{\eta}_{{k}_{1}} \hat{\eta}^{\dagger}_{{k}_{2}} >
<\hat{\eta}_{{k}_{2}} \hat{\eta}^{\dagger}_{{k}_{1}} >W_{2} + \]
\[ W_{2} <\hat{\eta}_{{k}_{1}} \hat{\rho}^{\dagger}_{{k}_{2}} >
<\hat{\rho}_{{k}_{2}} \hat{\eta}^{\dagger}_{{k}_{1}} >W_{1} + \]
\[ W_{1} <\hat{\rho}_{{k}_{1}} \hat{\eta}^{\dagger}_{{k}_{2}} >
<\hat{\eta}_{{k}_{2}} \hat{\rho}^{\dagger}_{{k}_{1}} >W_{2}. \]

Making the trace and limite,
the total expression for $n=2$ can be represented symbolicaly as:
\[ i \int^{t}_{0} dt_{1}\frac{i}{2} \int^{{t}_{1}}_{0}dt_{2}
W_{j}<\hat{\rho}_{j}(t_{1})\hat{\rho}_{l}(t_{2})>
<\hat{\rho}_{l}(t_{2})\hat{\rho}_{j}(t_{1})>W_{l}, \]
which admits the diagrammatic representation of fig 1a. where
$j,l = 1,2$ and $\hat{\rho}_{1}$ is representing $\hat{\rho}$
and $\hat{\rho}_{2}$ $\hat{\eta}$. \\

For $n=3$, one gets similarly:

 \[ i\int^{t}_{0}dt_{1} i\int^{{t}_{1}}_{0}dt_{2}
\frac{i}{3}\int^{{t}_{2}}_{0}dt_{3}W_{j}W_{l}W_{m}
<\hat{\rho}_{j}(t_{1})\hat{\rho}_{l}(t_{2})>
<\hat{\rho}_{l}(t_{2})\hat{\rho}_{m}(t_{3})>
<\hat{\rho}_{m}(t_{3})\hat{\rho}_{j}(t_{1})> \]
for which the diagramatic representation is given in fig 1b. \\

This procedure is easily extended to order $n$, allowing one to cast
the right-hand of (64) in
\beq
i \int^{t}_{0} dt_{1} <\Psi \mid \bar{\hat{\Gamma}} (t_{1},0)\initi,
\eeq
where $\bar{\hat{\Gamma}}(t_{1},0) $ is given by eq. (44). \\ \\

\appendix{Appendix B}

To calculate the fourth order contribution to the evolution operator
expansion, we initially separate the gaussian terms of the sum in the
exponential factor of eq (61), expanding this factor as
\beq
\final \hat{U}(t,0) \initi \\
\limite \left [ N e^{iS^{(0)}} \int D[\Phi] exp \left \{
i \frac{\epsilon}{2} \sum_{k_{1},k_{2}} \Phi_{k_{1}} S^{(2)}
\Phi_{k_{2}} \right \} (1 + i S_{3} + .....) \right ]
\eeq
where $S_{3}$ is the short notation for
\[ {\sum}^{\infty}_{n=3} \frac{1}{n!} \left(
\frac{ \partial^{n}S_{\Phi}}
{\partial \Phi^{n}} \right)_{(\xi = \xi^{*} = 0)} \Phi^{n}. \]

As the above expression involves the gaussian factor and the
polynomial expression, the lowest nonvanishing contribution appear for
$n=4$. To calculate (63) we use the generating function,
$Z(J^{*}_{j}, J_{j})$, defined as

\[ Z(J^{*}_{j},J_{j})) =  \int D[x^{*},x] exp \left\{
-x^{*}_{j}H_{jl}x_{l} + J^{*}_{j}x_{j} + J_{j}x^{*}_{j} \right \} \]
to the fourth order one gets
\[ {\final} \hat{U}(t,0) \initi = \limite \final\hat{\it{U}}^{(0)}(t,0)
\initi (1+ iS^{(4)} ) \]
with
\[ S^{(4)} = \frac{1}{6} \sum_{k_{1},k_{2}} \frac{\epsilon^{2}}{q^{2}}
<\hat{H}_{int}(t_{{k}_{1}}) \hat{H}_{int}(t_{{k}_{2}})>_{linked} \]

\newpage

\vspace*{10cm}
fig. 1- Diagrammatic representation of contribution to the exponential
factor in eq. (41). The indices $j,l,m,n , etc = 1,2$ are representing
the two  differents trial interaction  $W_{1}$ and $W_{2}$. \\

\vspace*{10cm}
fig. 2- Diagrammatic representation of contribution to $S^{(4)}$ in eq.
(45). The indices, $j,l = 1,2$ are representing the two-body potential
interaction from  subsystem $1$ $V_{1}$ and $2$ $V_{2}$.
\end{document}

----------------------------- End of body part 2